\documentclass[prb,preprint,amsmath,amssymb,superscriptaddress,notitlepage]{revtex4-2}
\usepackage{epsfig}
\usepackage{graphicx}
\usepackage{color}
\usepackage{bm}
\usepackage{svg}
\usepackage{upgreek}
\usepackage{soul}

\begin{document}

\title{Altermagnetism imaged and controlled down to the nanoscale}
\author{O.~J.~Amin}
\affiliation{School of Physics and Astronomy, University of Nottingham, Nottingham, UK}

\author{A.~Dal~Din}
\affiliation{School of Physics and Astronomy, University of Nottingham, Nottingham, UK}

\author{E.~Golias}
\affiliation{MAX IV Laboratory, Lund, Sweden}

\author{Y.~Niu}
\affiliation{MAX IV Laboratory, Lund, Sweden}

\author{A.~Zakharov}
\affiliation{MAX IV Laboratory, Lund, Sweden}

\author{S.~C.~Fromage}
\affiliation{School of Physics and Astronomy, University of Nottingham, Nottingham, UK}

\author{C.~J.~B.~Fields}
\affiliation{School of Physics and Astronomy, University of Nottingham, Nottingham, UK}

\author{S.~L.~Heywood}
\affiliation{School of Physics and Astronomy, University of Nottingham, Nottingham, UK}

\author{R.~B.~Cousins}
\affiliation{Nanoscale and Microscale Research Centre, University of Nottingham, Nottingham, UK}

\author{J.~Krempasky}
\affiliation{Photon Science Division, Paul Scherrer Institut, Villigen, Switzerland}

\author{J.~H.~Dil}
\affiliation{Photon Science Division, Paul Scherrer Institut, Villigen, Switzerland}
\affiliation{Institut de Physique, \'Ecole Polytechnique F\'ed\'erale de Lausanne, Lausanne, Switzerland}

\author{D.~Kriegner}
\affiliation{Institute of Physics, Czech Academy of Sciences, Prague, Czech Republic}

\author{B.~Kiraly}
\affiliation{School of Physics and Astronomy, University of Nottingham, Nottingham, UK}

\author{R.~P.~Campion}
\affiliation{School of Physics and Astronomy, University of Nottingham, Nottingham, UK}

\author{A.~W.~Rushforth}
\affiliation{School of Physics and Astronomy, University of Nottingham, Nottingham, UK}

\author{K.~W.~Edmonds}
\affiliation{School of Physics and Astronomy, University of Nottingham, Nottingham, UK}

\author{S.~S.~Dhesi}
\affiliation{Diamond Light Source, Harwell Science \& Innovation Campus, Didcot, UK}

\author{L.~\v{S}mejkal}
\affiliation{Institute of Physics, Czech Academy of Sciences, Prague, Czech Republic}
\affiliation{Institute of Physics, Johannes Gutenberg University, Mainz, Germany}

\author{T.~Jungwirth}
\affiliation{School of Physics and Astronomy, University of Nottingham, Nottingham, UK}
\affiliation{Institute of Physics, Czech Academy of Sciences, Prague, Czech Republic}

\author{P.~Wadley}
\affiliation{School of Physics and Astronomy, University of Nottingham, Nottingham, UK}

\maketitle

{\bf 
Nanoscale detection and control of the magnetic order underpins a broad spectrum of fundamental research and practical device applications. The key principle involved is the breaking of time-reversal ($\cal{T}$) symmetry, which in ferromagnets is generated by an internal magnetization. However, the presence of a net-magnetization also imposes severe limitations on compatibility with other prominent phases ranging from superconductors to topological insulators, as well as on spintronic device scalability. Recently, altermagnetism has been proposed as a solution to this restriction, since it shares the enabling ($\cal{T}$)-symmetry breaking characteristic of ferromagnetism, combined with the antiferromagnetic-like vanishing net-magnetization \cite{Smejkal2020,Smejkal2021a,Smejkal2022a}. To date, altermagnetic ordering has been inferred from spatially averaged probes \cite{Smejkal2022AHEReview,Feng2022,Reichlova2024,Betancourt2021,Tschirner2023,Kluczyk2023,Wang2023a,Han2024,Hariki2023,Fedchenko2024,Krempasky2024,Lee2024a,Osumi2024,Reimers2024,Hajlaoui2024,Lin2024}. Here, we demonstrate nanoscale imaging and control of altermagnetic ordering ranging from nanoscale vortices to domain walls to microscale single-domain states in MnTe \cite{Smejkal2021a,Betancourt2021,Lovesey2023b, Mazin2023,Krempasky2024,Lee2024a,Osumi2024,Hajlaoui2024,Kluczyk2023}. We combine the ($\cal{T}$)-symmetry breaking sensitivity of X-ray magnetic circular dichroism \cite{Hariki2023} with magnetic linear dichroism and photoemission electron microscopy, to achieve detailed imaging of the local altermagnetic ordering vector. A rich variety of spin configurations can be imposed using microstructure patterning or thermal cycling in magnetic fields. The demonstrated detection and control of altermagnetism paves the way for future research ranging from ultra-scalable digital and neuromorphic spintronic devices, to the interplay of altermagnetism with non-dissipative superconducting or topological phases.
} 

\newpage

For condensed matter physics, the $d$-wave (or higher even-parity wave) spin-polarization order in altermagnets represents the sought-after, but for many decades elusive, counterpart in magnetism of the unconventional $d$-wave order parameter in high-$T_c$ superconductivity \cite{Smejkal2022a}. For spintronics, altermagnets can merge favorable characteristics of conventional ferromagnets and antiferromagnets, considered for a century as mutually exclusive \cite{Smejkal2022a}. They can combine strong spin-current effects, which underpins reading and writing functionalities in commercial ferromagnetic memory bits, with vanishing net magnetization, enabling demonstrations of high spatial, temporal and energy scalability in experimental antiferromagnetic bits insensitive to external magnetic-field perturbations. These examples, underlined by the abundance of robust altermagnetism among materials ranging from insulators and semiconductors to metals and superconductors, illustrate the expected broad impact of altermagnets on modern science and technology \cite{Smejkal2022a}. 

To date, however, the unconventional properties of altermagnets have been experimentally detected using only spatially-averaging electronic transport \cite{Smejkal2022AHEReview,Feng2022,Reichlova2024,Betancourt2021,Tschirner2023,Kluczyk2023,Wang2023a,Han2024} or spectroscopy probes \cite{Hariki2023,Fedchenko2024,Krempasky2024,Lee2024a,Osumi2024,Reimers2024,Hajlaoui2024,Lin2024}. In our work, we report full-vector mapping of the altermagnetic order-vector and demonstrate the control, from nano- to microscale, of a rich landscape of altermagnetic textures, including vortices, domain walls and domains. We employ polarized X-ray photoemission electron microscopy (PEEM) which is a powerful real-space imaging technique in magnetism, offering the optimal combination of microscale field of view and nanoscale detection limit.

The measurements are performed at 100~K on a 30~nm thick film of $\alpha$-MnTe(0001) deposited on an InP(111)A substrate. MnTe is one of the prototypical materials in altermagnetic research \cite{Smejkal2021a,Betancourt2021,Hariki2023,Lovesey2023b,Krempasky2024,Lee2024a,Osumi2024,Hajlaoui2024,Kluczyk2023}. Below the transition temperature of 310~K, it has two Mn atoms in the unit cell carrying magnetic moments ${\bf M}_1$ and ${\bf M}_2$ of equal magnitude and opposite direction in the (0001)-plane, as shown in Fig.~1a. The two MnTe sublattices containing the opposite magnetic moments are connected by a spin symmetry combining a spin-space two-fold rotation with a real-space non-symmorphic sixfold screw-axis rotation ($[C_2\parallel C_6{\bf t}_{1/2}]$), and not with translation or inversion \cite{Smejkal2021a,Betancourt2021}. This non-relativistic spin-symmetry of the crystal structure generates an altermagnetic ($g$-wave) spin polarization which breaks the $\cal{T}$-symmetry of the electronic structure \cite{Smejkal2021a}. The perturbative relativistic spin-orbit coupling generates a weak magnetization along the [0001] axis which, in zero external magnetic field, only reaches a scale of $10^{-3}$~$\mu_{\rm B}$ per Mn (see Supplementary material) \cite{Smejkal2021a,Kluczyk2023,Hariki2023}. 

Our full-vector mapping includes the local real-space detection of the orientation of the altermagnetic order vector, ${\bf L}={\bf M}_1-{\bf M}_2$, with respect to the MnTe crystal axes by X-ray magnetic linear dichroism (XMLD) PEEM, and of the sign of {\bf L} for a given crystal-orientation by including X-ray magnetic circular dichroism (XMCD) PEEM. In antiferromagnets with opposite spin-sublattices connected by translation or inversion, the $\cal{T}$-odd XMCD is excluded by symmetry. In such cases, only the {\bf L}-axis can be detected by the $\cal{T}$-even XMLD-PEEM, but the sign of {\bf L} remains unresolved \cite{Nolting2000,Wadley2018,Chmiel2018,Krizek2020a,Jani2021,Amin2022}. Contrary to this, the recent theoretical and experimental spectroscopic study of altermagnetic MnTe has demonstrated the presence of a sizable XMCD, reflecting the $\cal{T}$-symmetry breaking in the electronic structure by the altermagnetic $g$-wave spin polarization \cite{Hariki2023}. Furthermore, the XMCD spectral shape due to {\bf L} pointing in the (0001)-plane is qualitatively distinct from the XMCD spectral shape due to a net magnetization $\bf{M}=\bf{M}_{1}+\bf{M}_{2}$ along the [0001]-axis \cite{Hariki2023}. We perform normal incidence XMCD-PEEM imaging at zero external field, where the weak relativistic remnant {\bf M} generates a negligible XMCD signal, and the detected spectral shape that gives rise to our measured XMCD-PEEM contrast is due to ${\bf L}\parallel\langle1\bar{1}00\rangle$ directions in the (0001)-plane, consistent with theoretical prediction \cite{Hariki2023}.

The method of combining the XMCD-PEEM and XMLD-PEEM images into the full-vector map of {\bf L} is illustrated in Fig.~1b. As the {\bf L}-vector subtends the angle, $\phi$, in the MnTe (0001)-plane relative to the [1$\bar{1}$00] axis, the $\text{XMCD}\propto\cos(3\phi)$, with maximum magnitude for ${\bf L}\parallel\langle1\bar{1}00\rangle$-axes and vanishing for ${\bf L}\parallel\langle2\bar{1}\bar{1}0\rangle$-axes \cite{Hariki2023}. An XMCD-PEEM image of a 25~$\mu$m$^2$ unpatterned area of MnTe is shown in Fig.~1c, where positive and negative XMCD appear as light and dark contrast, respectively. The corresponding three-colour XMLD-PEEM map, shown in Fig.~1d, was obtained from a set of PEEM images taken with the X-ray linear polarization rotated, within the MnTe (0001)-plane, in $10^{\circ}$ steps from $-90^{\circ}$ to $+90^{\circ}$ relative to the horizontal, [$1\bar{1}00$] axis. In this image, the local {\bf L}-vector axis is distinguished (by red-green-blue colours), but the absolute direction remains unresolved. This information is included by combining the XMCD-PEEM and XMLD-PEEM in a six-colour full-vector map, shown in Figs.~1e,f, where positive XMCD regions change the colour (red-green-blue to orange-yellow-purple) of the XMLD-PEEM map and negative XMCD regions leave it unchanged.

The characteristic full-vector mapping of {\bf L} in our unpatterned MnTe film, shown in Figs. 1e,f, displays a rich landscape of (meta)stable textures akin to earlier reports in compensated magnets \cite{Wadley2018,Chmiel2018,Krizek2020a,Jani2021,Amin2022}. There exist $60^{\circ}$ and $120^{\circ}$ domain walls separating domains with {\bf L} aligned along the different easy-axes, as well as vortex-like textures. Highlighted in Fig.~1f is an example of an altermagnetic vortex-antivortex pair, analogous to magnetic textures previously detected in antiferromagnets such as CuMnAs \cite{Amin2022}. However, only the XMLD-PEEM was available in the antiferromagnet \cite{Amin2022}, i.e., only the spatially varying N\'eel-vector axis could be identified, similar to our XMLD-PEEM image in Fig.~1d. In our altermagnetic case, we can add the information from the measured XMCD-PEEM (Fig.~1c). This allows us to plot the full-vector map of {\bf L}, as shown in Figs.~1e,f. We directly experimentally determine that the {\bf L}-vector makes a clockwise rotation by $360^\circ$ around the first vortex nano-texture, indicated by the magenta-white circle, while the other nano-texture is an antivortex with an opposite winding of the {\bf L}-vector, indicated by the cyan-white circle.

In Fig.~2 we demonstrate a designed formation and imaging of a vortex pair with predetermined winding and position. We utilize the combined effects of magnetoelastic forces and surface anisotropy which can result in alignment of the $\bf{L}$-vector with respect to a patterned edge of a compensated magnet \cite{Gomonay2014, Folven2011, Reimers2022, Reimers2023c}. We leverage this by patterning, using electron beam lithography and argon ion milling, a MnTe structure of a filled hexagon shape with edges along the $\langle1\bar{1}00\rangle$ easy-axes. In a virgin state, the interior of the hexagon splits into six wedge-shape domains with the {\bf L}-vector axes aligned parallel to the hexagon edges, and with domain walls extending from the hexagon corners towards the center of the structure (see Fig.~2b,c). Two domains from opposite edges of the hexagon can have their {\bf L}-vectors parallel (one pair in Fig.~2b,c), or antiparallel (two pairs in Fig.~2b,c). In the next step, we select one sign of the {\bf L}-vector in each domain pair by first warming the structure above the MnTe magnetic transition temperature, and then cooling it back to $100$~K in an external magnetic field of 0.4~T applied along the [0001] axis. In agreement with earlier spatially-averaging measurements of the anomalous Hall effect and XMCD spectra, and explained by the coupling of the external field to {\bf M} and of {\bf M} to {\bf L} \cite{Betancourt2021,Hariki2023}, this procedure results in the population of only one sign of {\bf L} in each pair of the $\langle1\bar{1}00\rangle$ easy-axis domains (see Fig.~2d,e). The formation of a vortex pair in the centre of the structure is then required to resolve the total winding angle of the {\bf L}-vector through $720^\circ$. In Figs.~2f-i we show analogous measurements in a larger hexagon. The observed magnetic configurations in the virgin state and after field cooling are similar to those in Figs.~2b-e near the hexagon edges, while in the central region they contain more complex textures reminiscent of the unpatterned film from Fig.~1. This highlights the important role of both the shape and scale of the patterned microstructures in the control of the magnetic configurations of altermagnetic MnTe.

Moving from the nanoscale vortices to the opposite, large-scale limit of the real-space control and detection of the altermagnetic states, we demonstrate in Fig.~3 a designed formation of single-domain states in MnTe. Here we focus on a patterned unfilled hexagon shape with 10~$\mu$m arm-length and 2~$\mu$m arm-width and arms along the $\langle1\bar{1}00\rangle$ easy-axes. In the virgin state, the patterning alone generates large domain states with the axis of the {\bf L}-vector determined by the crystal direction of the hexagon arm. This is seen in the XMLD-PEEM images in Fig.~3a. The arms also display narrow $180^\circ$ domain-wall lines with opposite contrast to the domains. In Fig.~3b we show the XMCD-PEEM image of the hexagon and in Fig.~3c the full vector map obtained from the combined XMCD and XMLD-PEEM images (see Supplementary material). Regions within the hexagon arms where the XMCD-PEEM contrast reverses confirm the presence of $180^\circ$ domain walls separating opposite {\bf L}-vector domains. Similarly, at the corners of the hexagon, XMCD-PEEM contrast reversal indicates $60^\circ$ domain walls separating the {\bf L}-vector domains in adjacent arms, and no contrast reversal indicates $120^\circ$ domain walls. To turn each arm into a micron-scale single-domain state, we apply the field cooling procedure as in Fig.~2. The removal of the domain walls and the formation of the single-domain states in the arms is shown in the XMCD-PEEM image and full vector map in Fig.~3d,e, respectively. In Fig.~3f,g we demonstrate that reversing the direction of the magnetic field applied during cooling results in a reversal of the direction of $\bf L$ in each of the single domain states.

The full-vector imaging and control of altermagnetic configurations ranging from nanoscale vortices to domain walls and large-scale domains, demonstrated in this work, has broad science and technology implications. It is the basis on which the new experimental field can develop, leveraging the ${\cal{T}}$-symmetry breaking phenomenology, vanishing magnetization, ultrafast dynamics, and compatibility of the altermagnetic order with the full range of conduction types from insulators to superconductors \cite{Smejkal2022a}. The ability to image and control micron-scale single-domain states will be highly relevant in the research of topologically protected Chern-insulator or Majorana states. Apart from these prospective non-dissipative and quantum-computing technologies, the same applies to digital spintronics where altermagnets have the potential to break spatial, temporal and energy scalability limits of the present ferromagnetic technology. Multidomain states with spatially varying magnetic configurations represent a complementary area which can leverage the unique phenomenology of altermagnets in the research of topological skyrmions, merons and other magnetic textures, and in the related field of neuromorphic devices. Our demonstration of the full-vector mapping and control of the magnetic-order in the nanoscale altermagnetic textures opens this experimental research front.

\newpage

\noindent{\bf References}
\bibliographystyle{naturemag}
\bibliography{Refs}

\protect\newpage

\noindent{\large\bf Methods}

\noindent{\bf Sample fabrication}

The 30~nm $\alpha$-MnTe films used for this study were grown at $\sim700$~K by molecular beam epitaxy (MBE) on InP(111)A substrates. The MnTe $c$-axis is oriented parallel to the normal of the substrate surface. We confirm the correct crystallographic phase and growth orientation of our MnTe films using X-ray diffraction, shown in the Supplementary material. Sample A (Fig.~1) was an uncapped $\alpha$-MnTe film kept under ultrahigh vacuum conditions and transported between the MBE and the PEEM in a custom-built vacuum suitcase. Sample B (Fig.~2 and 3) was an $\alpha$-MnTe film capped with 2~nm of Al to prevent surface oxidation of the MnTe layer. We carried out microfabrication on Sample B by coating with a 200~nm layer of PMMA then exposing
via EBL and developed in MIBK:IPA. Ar ion milling was used to fully remove the MnTe layer in the exposed areas before any residual resist was removed in acetone. using a combination of electron-beam lithography and argon ion milling.

\noindent{\bf PEEM imaging and N\'eel vector mapping}

The X-ray PEEM measurements were performed at the MAXPEEM beamline of the MAX IV Laboratory synchrotron. The X-ray beam was incident normal to the sample surface, with X-ray linear polarization vector in-plane and helicity vector out-of-plane. The linear dichroism asymmetry, $\text{XMLD}=(I(E_{1})-I(E_{2}))/(I(E_{1})+I(E_{2}))$, was calculated between images obtained at energies, $E_1$ and $E_2$, which correspond to maximum and minimum points in the magnetic linear dichroism spectra at the Mn $L_3$ absorption peak. The circular dichroism asymmetry, $\text{XMCD}=(I(\mu_{+})-I(\mu_{-}))/(I(\mu_{+})+I(\mu_{-}))$, was calculated between images obtained with opposite helicity polarizations, $\mu_{\pm}$, for a fixed energy corresponding to a maximum in the magnetic circular dichroism at the Mn $L_2$ absorption peak.

XMLD maps were produced from dichroism asymmetry images with X-ray linear polarization at angles, $\theta = -90^{\circ}$ to $90^{\circ}$, relative to the horizontal axis, in steps of $10^{\circ}$. The angular dependence of the XMLD was fitted with a $\sin(2(\theta+\varphi))$ function, where the phase offset, $\varphi$, encodes information about the local N\'eel vector axis. The symmetry along the axis is broken by the XMCD, which is used as a mask to produce the full vector maps. Extended details of the vector mapping process are included in the Supplementary material.

\noindent{\bf Field cooling}

Preparation of the sample in a field cooled state was done \textit{in situ} within the PEEM chamber. A permanent magnet was held in close proximity to the sample, with a 0.4~T field along the normal direction to the sample surface. The sample was heated to $350$~K, above the magnetic transition temperature of $310$~K. Liquid nitrogen cooling was used to reduce the sample temperature, through the magnetic transition, to a base temperature of $100$~K. For more details about the field cool setup, see the Supplementary material.

\protect\newpage

\noindent{\bf Acknowledgement}

We thank MAX IV Laboratory for time on Beamline MaxPEEM under proposal 20231714 (OJA). Research conducted at MAX IV, a Swedish national user facility, is supported by the Swedish Research Council under contract 2018-07152, the Swedish Governmental Agency for Innovation Systems under contract 2018-04969, and Formas under contract 2019-02496. OJA acknowledges support from the Leverhulme Trust Grant ECF-2023-755. The work was supported by the EPSRC Grant EP/V031201/1 and Altermag Grant 101095925.

\noindent{\bf Author contributions}

OJA, ADD, KWE, LS, TJ, and PW conceived and led the project. OJA, ADD, RPC, SLH, RBC, and AWR contributed to growth and fabrication of materials and devices. OJA, ADD, PW, KWE, BK, CJBF, SCF, EG, YN, AZ, SSD performed the XPEEM experiments and data analysis. OJA, ADD, PW, KWE, CJBF, SCF, SSD, DK, JK, and JHD performed sample characterisation. PW, TJ, SSD, OJA, ADD, and KWE wrote the manuscript with feedback from all authors.

\noindent{\bf Competing interests}

The authors declare no competing interests.

\noindent{\bf Data availability}

The data supporting the findings of this study are available from the corresponding author upon reasonable request.

\protect\newpage

\begin{figure}[h!]
\hspace*{0cm}\epsfig{width=1\columnwidth,angle=0,file=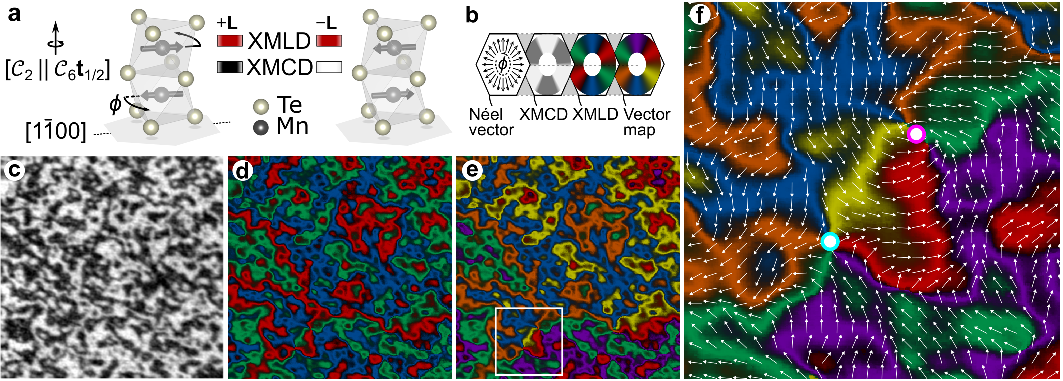}
\caption{\footnotesize\textbf{Mapping of the altermagnetic order vector in MnTe.}
\textbf{a,} Unit cell of $\alpha$-MnTe with a spin-space two-fold rotation and real-space non-symmorphic sixfold screw-axis rotation symmetry connecting the opposite-spin sublattices. The [1$\bar{1}$00] magnetic easy axis is highlighted with Mn spins collinear. Applying $T$ transforms the left unit cell into the right. Both unit cells with opposite N\'eel vector produce the same XMLD, but inequivalent XMCD signals due to $T$-symmetry breaking in the electronic structure of altermagnetic MnTe.
\textbf{b,} Illustration of the full vector mapping process. The colour wheels show the angular dependence of the XMCD, three-colour XMLD and six-colour full vector map on the in-plane {\bf L}-vector direction. The XMCD acts on the three-colour XMLD, with light XMCD regions changing the colour and dark XMCD regions leaving it unchanged to produce the six-colour {\bf L}-vector map. In the XMLD and full vector map, coloured segments indicate the magnetic easy axes oriented along the $\langle 1\bar{1}00\rangle$ crystallographic directions.
\textbf{c-e,} XMCD-PEEM and XMLD-PEEM, and full vector map respectively, of a 25~$\mu$m$^2$ region of unpatterned MnTe film.
\textbf{f,} An expanded view of a $\sim2$~$\mu$m$^2$ region in which a vortex-antivortex pair is identified. The vortex-antivortex core positions are highlighted by the magenta-white and cyan-white circles, respectively. The combination of XMLD-PEEM and XMCD-PEEM imaging allows for unambiguous determination of the helicity of the swirling textures of the altermagnetic order vector, indicated by the six colours and overlaid vector plot.
}
\label{f1}

\end{figure}

\begin{figure}[h!]
\hspace*{0cm}\epsfig{width=1\columnwidth,angle=0,file=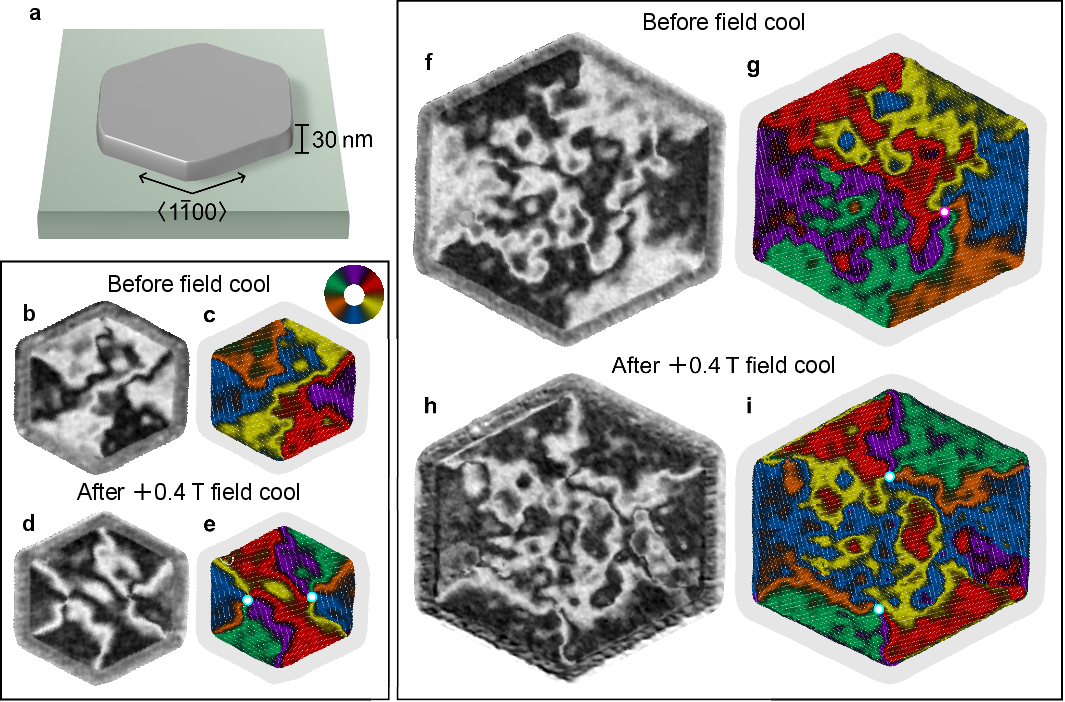}
\caption{\textbf{Controlled formation of altermagnetic vortex nano-textures.} 
\textbf{a,} Hexagon microstructure patterned in MnTe with edges along the $\langle1\bar{1}00\rangle$ easy-axes. \textbf{b, c,} A light/dark XMCD-PEEM map and a six-colour vector map, respectively, of the virgin state of a 2~$\mu$m wide filled hexagon. The {\bf L}-vector axis preferentially aligns parallel to the hexagon edges with domain walls forming at the hexagon corners.
\textbf{d, e,} Same as b, c, after field cooling with a 0.4~T field applied along the [0001]-axis. The {\bf L}-vector is restricted to a single direction for each axis parallel to the hexagon edges. This results in the formation of only three domain types with $120^\circ$ domain walls separating them at the hexagon corners, and of a vortex pair in the center of the structure, with core positions indicated by cyan-white circles. \textbf{f-i,} Same as b-e for a 4~$\mu$m wide hexagon. Near the edges, the sample behaves similarly to the smaller hexagon. In the central region it shows more complex magnetic textures reminiscent of the unpatterned film from Fig.~1.
}
\label{f2}
\end{figure}

\begin{figure}[h!]
\hspace*{0cm}\epsfig{width=1\columnwidth,angle=0,file=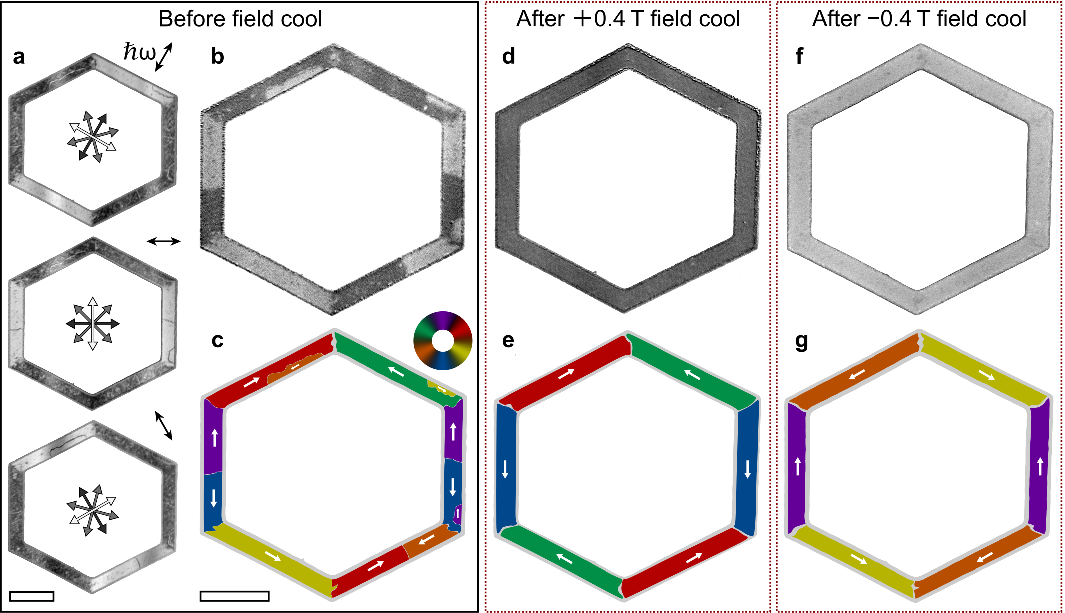}
\caption{\textbf{Large single-domain altermagnetic states controlled by micropatterning and field cooling.} 
{\bf Left, middle, and right panels} show images of an unfilled hexagon shape with arms, of 10~$\mu$m length and 2~$\mu$m width, aligned along the $\langle1\bar{1}00\rangle$ easy-axes, before field cooling and after field cooling with $+0.4$~T and $-0.4$~T, respectively.
{\bf a,} XMLD-PEEM images of the hexagon before field cooling for three directions of the X-ray linear polarization, indicated by the double-headed arrow in the top right corner of each image. The XMLD-PEEM contrast (double-headed arrows at the centre of each image) appears as light when {\bf L} is perpendicular to the X-ray polarization, indicating large single spin axis domains in each arm, parallel to the arm edge. $180^{\circ}$ domain walls can be seen as thin, contrasting lines, separating domains with opposite direction of {\bf L}. Both spatial scale bars correspond to 5~$\mu$m.
{\bf b,} The corresponding XMCD-PEEM image reveals the direction of {\bf L} along the spin axis parallel to the hexagon arms.
{\bf c,} A combination of the XMLD-PEEM and XMCD-PEEM images produces a six-colour vector map. 
{\bf d, e,} Repeats of b, c after field cooling the hexagon in a $+0.4$~T external magnetic field.
{\bf f, g,} Repeats of d, e after field cooling with the opposite sign magnetic field.
}
\label{f3}
\end{figure}

\end{document}